\def\papertitle{Flexible framework for audio reconstruction %\restoration% and quality enhancement
}
\def\paperauthorA{Ond\v{r}ej Mokr\'y}
\def\paperauthorB{Pavel Rajmic}
\def\paperauthorC{Pavel Z\'avi\v{s}ka}
\documentclass[twoside,a4paper]{article}
\usepackage{etoolbox}
\usepackage{dafx_20}
\usepackage{amsmath,amssymb,amsfonts,amsthm}
\usepackage{euscript}
\usepackage[utf8]{inputenc}
\usepackage[T1]{fontenc}
\usepackage{nimbusserif}
\usepackage{ifpdf}
\usepackage[english]{babel}
\usepackage{caption}
\usepackage{color}

\input glyphtounicode
\ninept

\newcounter{numauth}
\newcounter{listcnt}
\newcommand\authcnt[1]{\ifdefined#1 \stepcounter{numauth} \fi}

\newcommand\addauth[1]{
\ifdefined#1 
\stepcounter{listcnt}
\ifnum \value{listcnt}<\value{numauth}
\appto\authorslist{, #1}
\else
\appto\authorslist{~and~#1}
\fi
\fi}
\authcnt{\paperauthorB}
\authcnt{\paperauthorC}
\authcnt{\paperauthorD}
\authcnt{\paperauthorE}
\authcnt{\paperauthorF}
\authcnt{\paperauthorG}
\authcnt{\paperauthorH}
\authcnt{\paperauthorI}
\authcnt{\paperauthorJ}
\def\authorslist{\paperauthorA}
\addauth{\paperauthorB}
\addauth{\paperauthorC}
\addauth{\paperauthorD}
\addauth{\paperauthorE}
\addauth{\paperauthorF}
\addauth{\paperauthorG}
\addauth{\paperauthorH}
\addauth{\paperauthorI}
\addauth{\paperauthorJ}
\usepackage{times}
\newif\ifpdf
\ifx\pdfoutput\relax
\else
   \ifcase\pdfoutput
      \pdffalse
   \else
      \pdftrue
\fi

\ifpdf % compiling with pdflatex
  \usepackage[pdftex,
    pdftitle={\papertitle},
    pdfauthor={\authorslist},
    pdfsubject={Proceedings of the 23rd International Conference on Digital Audio Effects (DAFx-20)},
    colorlinks=false, % links are activated as color boxes instead of color text
    bookmarksnumbered, % use section numbers with bookmarks
    pdfstartview=XYZ, % start with zoom=100% instead of full screen; especially useful if working with a big screen :-)
  ]{hyperref}
  \usepackage[pdftex]{graphicx}
\else % compiling with latex
  \usepackage[dvips]{epsfig,graphicx}
  \usepackage[dvips,
    pdftitle={\papertitle},
    pdfauthor={\authorslist},
    pdfsubject={Proceedings of the 23rd International Conference on Digital Audio Effects (DAFx-20)},
    colorlinks=false, % no color links
    bookmarksnumbered, % use section numbers with bookmarks
    pdfstartview=XYZ % start with zoom=100% instead of full screen
  ]{hyperref}
\fi
\usepackage[hypcap=true]{caption}
\title{\papertitle}

\affiliation{
\paperauthorA, \paperauthorB \,\sthanks{Corresponding author.}and \paperauthorC}
{\href{http://splab.cz/en}{Signal Processing Laboratory} \\ Brno University of Technology, Brno, Czech Republic\\
{\tt \href{mailto:rajmic@feec.vutbr.cz}{rajmic@feec.vutbr.cz}}
}
\usepackage[linesnumbered,algoruled]{algorithm2e}
\usepackage[dvipsnames,table]{xcolor}

\SetCommentSty{mycommfont}

\usepackage{pgf}
\usepackage{tikz}
\usetikzlibrary{backgrounds,calc,positioning}

\usepackage{subcaption}

\def\RR{\mathbb{R}}

\def\CC{\mathbb{C}}

\def\NN{\mathbb{N}}

\theoremstyle{plain}

\theoremstyle{definition}

\theoremstyle{remark}

\DeclareTextAccent{\ring}{OT1}{23}

\def\x{\vect{x}}
\def\vv{\vect{v}}
\def\y{\vect{y}}
\def\z{\vect{z}}
\def\w{\vect{w}}

\def\sdr{\mathrm{SDR}}

\newcommand{\Id}{\mathrm{Id}}
\newcommand{\syn}{\adjoint{\ana}}
\newcommand{\ana}{A}

\newcommand{\norm}[1]{\|#1\|}
\newcommand{\abs}[1]{\left\vert#1\right\vert}

\newcommand{\adjoint}[1]{#1^*}

\newcommand{\vect}[1]{\mathbf{#1}} %vector
\newcommand{\matr}[1]{\mathbf{#1}} %matrix

\newcommand{\argmin}{\mathop{\operatorname{arg~min}}}

\newcommand{\prox}{\mathrm{prox}}

\newcommand{\proj}{\mathrm{proj}}

\newcommand{\GT}{\Gamma_\mathrm{T}}
\newcommand{\GTF}{\Gamma_\mathrm{TF}}
\newcommand{\Gone}{\Gamma_L}
\newcommand{\Gtwo}{\Gamma_K}
\newcommand{\signal}{\vect{y}}
\newcommand{\mask}{M}
\newcommand{\sparse}{\mathcal{S}}
\newcommand{\Tdim}{P}

\newcommand{\pT}{p_\mathrm{T}}
\newcommand{\bT}{b_\mathrm{T}}
\newcommand{\pTF}{p_\mathrm{TF}}
\newcommand{\bTF}{b_\mathrm{TF}}

\newcommand{\restoration}{reconstruction}
\begin{document}
% more pdf-tex settings:
\ifpdf % used graphic file format for pdflatex
  \DeclareGraphicsExtensions{.png,.jf,.pdf}
\else  % used graphic file format for latex
  \DeclareGraphicsExtensions{.eps}
\fi

%\makeatletter
%\pdfbookmark[0]{\@pdftitle}{title}
%\makeatother

\maketitle

\begin{abstract}

The paper presents a unified, flexible framework for the tasks of audio inpainting, declipping, and dequantization.
The concept is further extended to cover analogous degradation models in a~transformed domain, e.g.\ quantization of the signal's time-frequency coefficients.
The task of reconstructing an audio signal from degraded observations in two different domains is formulated as an inverse problem,
%regularized with a suitable functional.
and several algorithmic solutions are developed.
The viability of the presented concept is demonstrated on an example where
audio reconstruction from partial and quantized observations of both the time-domain signal and its time-frequency coefficients is carried out.

\end{abstract}

\section{Introduction}
\label{sec:introduction}

%\todo{Pane Kříži, řešíme, zda je lepší slovo reconstruction nebo restoration. Dle úsudku i dle Oxfordského slovníku se kloníme spíše k reconstruction, protože nejenom restaurujeme, ale skutečně vytváříme něco co předtím chybělo nebo bylo poničeno. V textu se tedy snažíme používat reconstruction.}

%\todo{PR: já bych souhlasil s nahrazením "restoration" pomocí "reconstruction" (rev.\,4). Pokud tak učiníme, tak při posledním čtení by to chtělo pohlídat ty pojmy i v textu. PZ: nemyslím si. OM: podle https://www.nps.gov/tps/standards/four-treatments.htm je to spíš reconstruction. PR: podle oxforďáka taky}

Audio inpainting, audio declipping and audio dequantization are \restoration{}\footnote{%
%\todo{We choose the terms \emph{to reconstruct/reconstruction} over \emph{to restore/restoration}.
%Such terminology well reflects the tasks of rebuilding the signal from the observed pieces.}%
%}
We choose the term \emph{reconstruction} over \emph{restoration}, as this
reflects well the task of rebuilding the signal from incomplete or degraded pieces.%
\rule[-8pt]{0pt}{0pt}%
}
tasks
that are usually studied separately in the literature.
In audio inpainting, some of the time-domain signal samples are completely missing
and they need to be recovered,
while in the cases of declipping and dequantization, the samples are not lost fully 
%entirely
and the samples to be recovered are known to lie in prescribed numerical ranges, depending on the
%exact 
model of the degradation.
The feasibility set is called consistent if any solution, when exposed to the considered degradation model, produces exactly the observed signal.
For example, in the case of audio inpainting, this shall be understood such that the reliable samples are kept intact.

A unification of different audio \restoration{} tasks
%and the possibility to combine several forms of degradation into a single formulation
has partially been discussed in 
\cite{RenckerBachWangPlumbley2018:Sparse.recovery.dictionary.learning},
where the authors covered dequantization and declipping (possibly at the same time),
and in
\cite{GaultierBertinKiticGribonval2017:Framework.audio.restoration,Gaultier2019:PhD.Thesis},
whose formulation allowed denoising and declipping (but not simultaneously). %at the same time).
A flexible algorithmic framework is also presented in \cite{Bilen2018:NTF_audio_inverse_problems}, based on the non-negative matrix factorization
(which is shown to be suitable for simultaneous audio declipping and click concealment).
The present article shows
%the way
how the three tasks can be covered by a~unified restoration framework,
all of them possibly taking effect at the same time.
The greatest contrast to the earlier attempts is, however, that
this paper extends the range of degradation models by additionally considering a transformed domain.
This is to say,
the missing, clipped and quantized observations are further allowed after (linearly)
transforming the signal, e.g.\ by the Short-time Fourier transform.

In Section \ref{sec:building.the.task}, we introduce the three respective audio degradation models in more detail,
emphasize their common factors,
and build a set of feasible time-domain signals, which contains the potential solutions to the recovery task.
We then extend the degradation to the transformed domain and present the synthesis and analysis variants of the resulting feasible set.

Finding a~solution to any of the described recovery tasks is generally ill-posed.
A~regularizer is needed to pick favorable candidates from the feasibility set.
The sparsity of time-frequency-transformed audio signals has been shown to be a suitable regularizer for audio recovery problems
\cite{Adler2012:Audio.inpainting,SiedenburgKowalskiDoerfler2014:Audio.declip.social.sparsity,MokryRajmic2019:Reweighted.l1.inpainting,ZaviskaRajmicSchimmel2019:Psychoacoustics.l1.declipping}.
Thus, Section \ref{sec:solving.the.task} presents a general optimization problem
with a~special emphasis on using the $\ell_1$ relaxation of true sparsity.
The section also presents a~single, unified algorithm
to find the numerical solution in the case of a convex regularizer.

In Section \ref{sec:experiment}, we present a proof-of-concept example
of an audio codec (i.e. coder and decoder).
In the coder part, the original, input audio signal is due to subsampling \emph{and}\/ quantization
in both the time \emph{and}\/ the time-frequency (TF) domains.
The decoder attempts to recover the signal from this partial information,
based on the assumption of sparsity of the (now unknown) original.
Today's audio codecs are built on the single-domain information,
for instance the classical MPEG model codes the TF coefficients only, based on the global masking threshold estimate
\cite{Shlien1994:Guide.to.MPEG-1}.
Recovery from quantized transformed observations is also studied in
\cite{JacquesHammondFadili2011:Dequantizing.Compressed.Sensing}
in the context of compressed sensing.
An interesting recent approach from 
\cite{Peter2019:Compressing.audio.inpainting.sparsificat}, which is inspired in the image processing field, subsamples and quantizes purely time-domain audio samples to achieve compression.
We show experimentally that in contrast to that approach,
splitting the available bit budget between the two domains can be beneficial in some cases.

%Regarding the notation, there will be no distinction between the closed and open intervals.
%It is clear that infinite values of the restored signal are unreasonable.

\section{Building the framework}
\label{sec:building.the.task}

\subsection{Time-frequency representations}

In audio processing, TF operators are usually used to provide a~suitable representation of a~signal \cite{Grochenig2001:Foundations.T-F.analysis}.
A~signal $\x\in\RR^P$ is represented as a superposition of time-localized oscillations, where the localization is due to the so-called window function that moves along the signal.
Among such TF operators, the so-called \emph{tight frames} are usually preferred, since they provide effective handling of both theoretical derivations and practical computations \cite{Grochenig2001:Foundations.T-F.analysis,Balazs2017:Frame.Theory.Psychoacoustics,ZaviskaRajmicPrusaVesely2018:RevisitingSSPADE}.
%
%Certain types of
The Short-time Fourier (STFT, also known as the Gabor transform) %\cite{christensen2014}
%\todo{jsi si jistý tou citací? já to neznám (PR)}
or the Modified Discrete Cosine (MDCT) transforms
\cite{Adler2012:Audio.inpainting,DerrienNecciariBalasz2015:ERB-MDCT}
are classical examples of such operators.

Throughout the paper, we use the following convention:
To obtain an expansion of a~signal $\x\in\RR^P$ to a~series of TF coefficients,
the \emph{analysis} operator $\ana\colon\RR^P\to\CC^Q$ is applied, where we assume $Q\geq P$.
Its adjoint, the \emph{synthesis} operator $\syn\colon\CC^Q\to\RR^P$\!, reproduces the time-domain signal from the coefficients.

A tight frame with frame bound $\alpha$ can be characterized by the property
$\syn\!\ana = \alpha\Id$,
where $\Id$ is the identity operator, here on the space $\RR^P$.
%Furthermore, denote $\RA\subset \CC^Q$ the range space of the analysis operator,
%i.e.\ the set of spectra consistent with the time-domain signals.
%In case of a tight frame with frame bound $\alpha$, the orthogonal projection onto $\RA$ is simply expressed as $\proj_{\RA} = \alpha^{-1}\ana\syn$ \cite{christensen2008}.
When the constant $\alpha = 1$, the frame is said to be Parseval tight.

\subsection{Inpainting}

Audio inpainting is a general term for recovering missing or highly degraded samples of the audio signal \cite{Adler2012:Audio.inpainting}.
Suppose $\signal\in\RR^\Tdim$ is the original, non-degraded signal, and $\mask\signal$ is the partial observation of $\signal$.
The operator $\mask\colon \RR^P \to \RR^P$ keeps the reliable samples, while putting zeros at the positions of missing or unreliable samples;
these positions are assumed to be known.
Thus, $\mask$ can be identified with a~diagonal matrix $\matr{M}$ of size $P\times P$,
for which $m_{pp} = 1$ for a~reliable sample $y_p$, $p=1,\ldots,P$, and zero otherwise.
The solution of the inpainting problem is % called \emph{consistent} if it lies in a~naturally defined set%
supposed to lie in a~naturally defined set
%
%\begin{equation}
$
	\GT^\mathrm{inp} = \left\{ \x\in\RR^\Tdim \mid \mask\x = \mask\signal \right\},
$
%\end{equation}
%
where the subscript indicates that it is defined in the time domain.
%The consistency shall be understood such that the observed samples are kept in the solution.

Clearly, defining the feasible set alone is not sufficient to solve the inpainting problem, since 
the inverse problem is ill-posed.
Thus, the path to a solution
must start from a careful consideration of additional assumptions about the signal.
To name but a few, the solution may be modeled as an autoregressive process \cite{javevr86,Etter1996:Interpolation_AR}, as a sum of sinusoidal components \cite{lagrange2005long}, or it is assumed to be sparse with respect to a~suitable TF transform
\cite{Adler2012:Audio.inpainting,MokryRajmic2019:Reweighted.l1.inpainting,MokryZaviskaRajmicVesely2019:SPAIN}.

For the purpose of further generalization, the time-domain set $\GT^\mathrm{inp}$ may be equivalently defined entrywise as% a box-type set%
\begin{equation}
	\x\in\GT^\mathrm{inp}\;\Leftrightarrow\;\begin{cases}
	&x_p \in [ y_p, y_p ] \ \text{for reliable indexes }p,\\
	%&x_p \in [ -\infty, +\infty ] \ \text{otherwise},
	&x_p \in ( -\infty, +\infty ) \ \text{otherwise}.
	\end{cases}
	\label{eq:gamma.inpainting}
\end{equation}
%
%for $p=1,2,\ldots,P$.
%\todo{závorky}
%\todo{je třeba zmínit, že "T" v indexu je "čas"?}

\subsection{Declipping}

Audio declipping aims at recovering a~signal damaged by clipping. 
This negative effect is one of the common audio degradation types and it can be described as a non-linear distortion causing a~limitation of a signal,
such that all values of the signal exceeding the allowed dynamic range defined by thresholds $[-\theta, \theta]$ are strictly limited by these thresholds. 
Because of the strict limitation of signal samples, the effect is also referred to as \emph{hard clipping}.
Not only does the information contained in the peaks get lost, but clipping also introduces a great number of higher harmonics, which leads to a significant reduction in the perceived audio quality \cite{Tan2003}
and also in the accuracy of automatic voice recognition
\cite{Malek2013:Blind.compensation}.

Audio declipping is similar to audio inpainting,
with the difference that in the case of audio declipping,
the additional information (lower or upper bounds) about the clipped samples is available. 
Simple inpainting methods are able to effectively perform declipping, such as the Janssen method used in \cite{Adler2012:Audio.inpainting}.
In general, however, inpainting approaches to declipping do not guarantee the consistency of the solution
with the clipping constraints.

Similarly to the inpainting case, the set of feasible solutions,
$\GT^\mathrm{dec}$,
is defined entrywise, taking advantage of the information that declipped samples need to exceed the clipping thresholds:
\begin{equation}
	\x\in\GT^\mathrm{dec}\Leftrightarrow
	\begin{cases}
	&\hspace{-.7em}x_p \in [ y_p, y_p ] \ \text{for reliable samples}\ y_p,\\
	&\hspace{-.7em}x_p \in ( {-}\infty, {-}\theta ] \ \text{for observed samples}\ {-}\theta,\\
	&\hspace{-.7em}x_p \in [ \theta, +\infty ) \ \text{for observed samples}\ \theta.
	\end{cases}
	\label{eq:gamma.declipping}
\end{equation}
%\todo{závorky}
%\todo{maskovací operátory nepotřebujeme?}

\subsection{Dequantization}

The term dequantization refers to an inverse problem where a signal should be recovered based on the knowledge of its quantized observation.
In this subsection, the quantization acts in the time domain, i.e.\ directly on the audio samples;
%In quantization,
%where the original sample
the original sample
is substituted with the value of the nearest quantization level.
The unique quantization level is identified using a~pair of the nearest so-called decision levels \cite{Gray1998:Quantization}.

More specifically, assume a~series of quantization levels
\begin{equation}
	\dots<q_{-1}<q_{0}<q_{1}<q_{2}<\dots,
	\label{eq:Quantiz.levels}
\end{equation}
where this sequence can be theoretically infinite (but is always finite in practice).
%Fix $p$ for the moment.
For a given $p$ and
%For the 
an input sample $y_p$ there exists a~unique $n$ such that it holds $q_n \leq y_p < q_{n+1}$.
Based on the decision level $d_n$, for which $q_n < d_{n} < q_{n+1}$,
quantization maps $y_p$ either to $q_n$ (when $y_p<d_n$)
or to $q_{n+1}$ (when $y_p\geq d_n$).
In turn, if a~quantized value $y_p^\mathrm{quant}$ is observed,
there exists a~single interval $[d_n,d_{n+1})$ to which $y_p$ belongs.

%All together,
Therefore, for the purpose of formulating a general problem,
the set of feasible solutions %consistent with the quantization model
is defined as the box-type set $\GT^\mathrm{deq}$,
\begin{equation}
	%\x\in\GT^\mathrm{deq}\;\Leftrightarrow\;x_p\in[\lambda_{n},\mu_{n}],
	\x\in\GT^\mathrm{deq}\;\Leftrightarrow\;x_p\in[d_n,d_{n+1}),
	\label{eq:gamma.dequantization}
\end{equation}
where $d_n$ and $d_{n+1}$ (the closest lower and the closest upper decision levels to $y_p$, respectively) change depending on $p$, which is intentionally not reflected by the notation.
% Here, $d_n$ and $d_{n+1}$ are the closest lower and the closest upper decision levels to $y_p$, respectively.
%is the interval the observed value $y_p$ belongs to.
%It should be emphasized that this interval depends on the sample $y_p$, which is not reflected by the notation in \eqref{eq:gamma.dequantization}.
%
Note also that in the finite case, border cases can be treated by using $\pm\infty$ in place of the lower or the upper bound in \eqref{eq:gamma.dequantization}.
In such a~case, the half-open interval should be replaced by an open interval.

\subsection{General formulation}

When working with digital signals, clipping can be seen as a special kind of quantization.
In such a case, the set 
%$\ldots\alpha_{-1}<\alpha_{0}<\alpha_{1}<\alpha_{2}\ldots$
of quantization levels defined by Eq.\,\eqref{eq:Quantiz.levels}
%covers precisely the range of possible values in the range $[-\theta, \theta]$.
corresponds exactly to the set of all possible numerical values in the range $[-\theta, \theta]$.

Looking at definitions \eqref{eq:gamma.inpainting}, \eqref{eq:gamma.declipping} and \eqref{eq:gamma.dequantization},
one may observe that it is straightforward to define a feasible set for \emph{simultaneous} audio inpainting, declipping and dequantization.
Such a~set is defined entrywise as a~multidimensional interval $\GT$
such that
\begin{equation}
	\x\in\GT\;\Leftrightarrow\;x_p\in[l_{\mathrm{T}p},u_{\mathrm{T}p}],\quad p = 1,2,\ldots,P,
	\label{eq:gamma.time}
\end{equation}
where the entries of the vector lower bound $\vect{l}_\mathrm{T}$ and the vector upper bound $\vect{u}_\mathrm{T}$
depend on the type of degradation that occurs at the index $p$, $p=1,\ldots,P$.
One can think of $\GT$ as a box in the $P$-dimensional space with its walls always parallel to an axis.
The bounds may formally be plus or minus infinity, and in such a~case, the box is infinitely wide in the respective directions.
%in case of inpainting, cf.\ \eqref{eq:gamma.inpainting}.

It is straightforward to show that the set $\GT$ is convex.
Furthermore, solving an inverse problem with such a set of feasible solutions is tractable since the projection onto this type of set is available explicitly and entrywise by
\begin{equation}
	\left(\proj_{\GT}(\x)\right)_p = \min\left\{ u_{\mathrm{T}p}, \max\left\{ x_p, l_{\mathrm{T}p} \right\} \right\}.
	\label{eq:projT}
\end{equation}

%\subsection{Using the information of two different domains}
\subsection{Feasible set in a transformed domain}

So far, only time-domain degradation has been considered,
leading to the set $\GT$.
Nevertheless, degradation
%of the same type
as presented above can also happen in a~transformed domain.
%(even in multiple domains).
The aim of this section is to generalize the above concept to both the time and the TF domains. % is taken into account.
%Such a~model is inspired by audio coders and decoders which use quantized TF coefficients of a time-domain signal.

Similarly to \eqref{eq:gamma.time},
we define a %\emph{consistent}\/
feasible set within the TF domain.
Such a~domain is generally a subset of $\CC^Q$.
Any interval shall be understood in such a way that the real and imaginary parts are considered independently.
As an example, for $l, u\in\CC$, we denote
\begin{equation}
	z\in[l,u]\;\Leftrightarrow\;\Re(z)\in[\Re(l),\Re(u)]\wedge	\Im(z)\in[\Im(l),\Im(u)].
\end{equation}
With such a notation, we define the membership in $\GTF$ as
%define the set
%
\begin{equation}
	\z\in\GTF\;\Leftrightarrow\;z_q\in[l_{\mathrm{TF}q},u_{\mathrm{TF}q}],\quad q = 1,2,\ldots,Q.
	\label{eq:gamma.time.frequency}
\end{equation}
The vector bounds $\vect{l}_\mathrm{TF},\vect{u}_\mathrm{TF}\in\CC^Q$
determine for each coefficient whether its clipped or quantized version is observed
or whether the coefficient is missing.

Combining the constraints in the two
%(or possibly even more)
domains reduces the size of the overall feasible set. % $\GT\cap\GTF$.
%which is valuable for finding the restored signal.
%This intersection is never empty, since it must contain at least the original, not degraded signal, 
In general, however, this is not enough, and additional prior information is necessary.

\subsection{Defining a prior}

For the purpose of the general framework, assume some knowledge about the TF coefficients invoked by minimizing a~functional $\sparse\circ W$.
As a~particular example, consider the (relaxed) sparse prior, namely % together with the $\ell_1$ relaxation.
%in such a~case,
$(\sparse\circ W)(\z) = \sparse(W\z) = \norm{W\z}_1$,
where $W$ is a~diagonal ope\-ra\-tor assigning weights to the respective coefficients.
The $\ell_1$ norm sums the magnitudes of the elements of its argument \cite{DonohoElad2003:Optimally}.

Combining the prior and the feasible sets $\GT$ and $\GTF$ provides us with the following general formulation:
\begin{equation}
 	\argmin_\vect{u}\ \sparse(W\!K\vect{u})\quad\text{subject to}\quad L\vect{u}\in\GT,\;K\vect{u}\in\GTF.
 	\label{eq:general.formulation}
\end{equation}
The linear operators $K$ and $L$ play the role of either the synthesis or the analysis operator of a~suitable TF transform.
In a~typical situation, one of them will be identity.
%\todo{Mluví se o typické situaci, ale není to spíš vynuceno, že jeden z nich musí být identita, jelikož jinak by to nesedělo s definicemi těch Gamma? Viz druhá červená...}
Such a notation may look redundant, % is that this way,
but the reason for this shape of the formulation \eqref{eq:general.formulation} is that it covers both the synthesis variant ($L=\syn$, $K=\Id$)
\begin{equation}
	\argmin_\z \ \sparse(W\z)\quad\text{subject to}\quad \syn\z\in\GT,\;\z\in\GTF,
	\label{eq:general.formulation.synthesis}
\end{equation}
%
%when $K$ is the identity, $K=\Id$,
and the analysis variant ($L=\Id$, $K=\ana$):
\begin{equation}
	\argmin_\x \ \sparse(W\!\ana\x)\quad\text{subject to}\quad \x\in\GT,\;\ana\x\in\GTF.
	\label{eq:general.formulation.analysis}
\end{equation}
Note that if a non-unitary transform $\ana$ is used,
the formulations \eqref{eq:general.formulation.synthesis} and \eqref{eq:general.formulation.analysis} are not equivalent.
Also, the feasible sets in Eq.\,\eqref{eq:general.formulation.synthesis} and \eqref{eq:general.formulation.analysis} may differ, as in the case of a~non-tight frame.

\section{Solving the task}
\label{sec:solving.the.task}

The important observation about the sets $\GT$ and $\GTF$ defined by \eqref{eq:gamma.time} and \eqref{eq:gamma.time.frequency}, respectively,
is that both are box-type (and thus convex) sets.
Furthermore, both the sets
$\Gone = \left\{\vect{u}\mid L\vect{u}\in\GT\right\}$
and
$\Gtwo = \left\{ \vect{u}\mid K\vect{u}\in\GTF\right\}$
are convex as well.
The reason is that the preimage
%(or inverse image)
of a convex set under a~linear operator is a~convex set, which is straightforward to show.
%
% source of terminology:
% https://en.wikipedia.org/wiki/Image_(mathematics)#Inverse_image
%
Finally, the intersection of two convex sets is once again a convex set, therefore the set of feasible solutions in the constrained formulation \eqref{eq:general.formulation} is convex
for arbitrary linear operators $L$ and $K$.

However, such an intersection is a~rather complicated set.
One of the sets $\Gone$ and $\Gtwo$ is no longer a~simple box-type set,
hence the intersection $\Gone\cap\Gtwo$ is generally a polyhedron either in the time domain (for the analysis model)
or in the TF domain (for the synthesis model).
%\todo{Tady mi něco nehraje; kdyby jeden z operátorů nebyl identita, formálně ten průnik ani nemůžeme udělat, protože jsou množiny obecně jiných dimenzí. \textbf{Stačí, aby začínaly ve stejném prostoru, prvky $\Gone$ i $\Gtwo$ jsou účka bez ohledu na to, kam je zobrazuje $L$ a $K$.}}
%This difficulty is treated right in the following section.
Still, it remains a non-empty set, since it must contain at least the original, non-degraded signal or coefficients. Thus, the formulation \eqref{eq:general.formulation} has a~solution.

%Based on the above observations, Sections \ref{sec:consistent.convex.1} and \ref{sec:consistent.convex.2} mainly focus on the convex setting.
%Sections \ref{sec:consistent.nonconvex}, \ref{sec:inconsistent.convex} and \ref{sec:inconsistent.nonconvex} will suggest alternative approaches;
%however, those will not be further developed in the present paper.

\begin{algorithm}[t]
	\footnotesize
	\KwIn{The linear operators $L_m,\;m=1,\dots, M$, the proximal operators $\prox_{h_m},\;m=1,\dots, M$, $\prox_{g}$ and the gradient $\nabla f$.}
	Choose the parameters $\tau,\sigma,\rho>0$.\\
	Choose the initial estimates $\vect{u}^{(0)},\vv^{(0)}_1,\dots,\vv^{(0)}_M$.\\
	\For{$i = 0,1,\dots$}{
		\For{$m = 1,\dots,M$}{
			$\tilde{\vv}_m^{(i+1)} = \prox_{\sigma \adjoint{h_m}}\left( \vv_m^{(i)} + \sigma L_m \vect{u}^{(i)}\right)$\\
			$\vv_m^{(i+1)} = \rho\tilde{\vv}_m^{(i+1)}+(1-\rho)\vv_m^{(i)}$\\
		}		
		$\tilde{\vect{u}}^{(i+1)} = \prox_{\tau g}\left( \vect{u}^{(i)} -\tau\nabla f\left(\vect{u}^{(i)}\right) - \tau\sum%_{m=1}^{M}
		\adjoint{L_m} \left(2\tilde{\vv}_m^{(i+1)} - \vv_m^{(i)}\right)\right)$\\
		$\vect{u}^{(i+1)} = \rho\tilde{\vect{u}}^{(i+1)} + (1-\rho)\vect{u}^{(i)}$
	}
	\KwOut{$\vect{u}^{(i+1)}$}
	\caption{The CV algorithm
		%no.\ 2
		for solving
		%the problem
		\eqref{eq:CV.problem}}
	\label{alg:CV.general}
\end{algorithm}

\subsection{Consistent convex approach, arbitrary linear operators}
\label{sec:consistent.convex.1}
We focus on the case when the function $\sparse$ is convex, thus the whole problem is convex.
Convexity implies that the there exists a single global minimum. % and that the (only) global minimum is achievable.
The idea is to use a~proximal splitting method \cite{combettes2011proximal} to solve the formulation \eqref{eq:general.formulation} numerically,
which allows us to focus separately on operations related to the function $\sparse$,
to the constraint $\vect{u}\in\Gone$
and to the constraint $\vect{u}\in\Gtwo$.
%\todo{Below, we employ the explicit projection of vectors onto a~box-type set.
%This is applicable in particular when the proximal operator of $\sparse$
%has an explicit form, which will be the case below.}

In the following, the notion of the proximal operator will be needed.
The proximal operator of a proper convex lower semi-continuous function $h\colon\mathbb{V}\to\RR$ %, denoted $\prox_h\colon\mathbb{V}\to\mathbb{V}$, is defined as
is a mapping from $\mathbb{V}$ to $\mathbb{V}$ defined at any point $\vect{u}\in\mathbb{V}$ by the minimization problem
$
	\prox_h(\vect{u}) = \argmin_\vv \left\{h(\vv) + \frac{1}{2}\norm{\vv-\vect{u}}^2\right\}.
$
Here, $\mathbb{V}$ stands for the Hilbert space $\RR^P$ or $\CC^Q$.
%To simplify the notation hereafter,

To design a~particular proximal algorithm,
the formulation \eqref{eq:general.formulation} is first rewritten into the unconstrained form 
using the so-called \emph{indicator function} $\iota_\Gamma$ of the set $\Gamma$.
For $\vect{u}\in\Gamma$, the function returns 0, and $\infty$ otherwise.
The formulation \eqref{eq:general.formulation} thus attains the form
\begin{equation}
	\argmin_\vect{u} \left\{ \sparse(W\!K\vect{u}) + \iota_{\GT}\left(L\vect{u}\right) + \iota_{\GTF}\left(K\vect{u}\right) \right\}.
	\label{eq:general.formulation.unconstrained}
\end{equation}
The unconstrained form is suitable for the use of the generic proximal algorithm proposed independently by Condat \cite{Condat2014:Generic.proximal.algorithm}
and Vũ 
\cite{Vu2011:splitting.dual.monotone.inclusions}
(further referred to as the CV algorithm).
It is tailored to solve problems of the form
\begin{equation}
	\argmin_\vect{u} \left\{ f(\vect{u}) + g(\vect{u}) + \sum_{m=1}^{M}h_m(L_m\vect{u}) \right\}, 
	\label{eq:CV.problem}
\end{equation}
where $f,g,h_1,\dots,h_m$ are convex lower semi-continuous functions, $f$ is differentiable, and $L_1,\dots,L_m$ are bounded linear operators.
We will utilize the second of the two proposed variants from \cite{Condat2014:Generic.proximal.algorithm},
the general form of which is reproduced in Alg.\,\ref{alg:CV.general}.

\begin{algorithm}[t]
	\footnotesize
	\KwIn{The linear operators $W,K,L$, the proximal operator $\prox_\sparse$ and the projectors $\proj_{\GT},\proj_{\GTF}$.}
	Choose the parameters $\tau,\sigma,\rho>0$ satisfying the conditions \eqref{eq:CV.convergence}.
	\\ Choose the initial estimates $\vect{u}^{(0)},\vv_1^{(0)},\vv_2^{(0)},\vv_3^{(0)}$.\\
	\For{$i = 0,1,\dots$}{
		\tcc{\scriptsize update corresponding to $h_1$}
		\vspace{0.1em}
		$\tilde{\vv}_1^{(i+1)} = \vv_1^{(i)} + \sigma WK \vect{u}^{(i)} - \sigma\prox_{\sparse/\sigma}\left( \vv_1^{(i)}/\sigma + WK \vect{u}^{(i)}\right)$\\
		$\vv_1^{(i+1)} = \rho\tilde{\vv}_1^{(i+1)}+(1-\rho)\vv_1^{(i)}$\\
		\vspace{0.1em}
		\tcc{\scriptsize update corresponding to $h_2$}
		\vspace{0.1em}
		$\tilde{\vv}_2^{(i+1)} = \vv_2^{(i)} + \sigma L \vect{u}^{(i)} - \sigma\proj_{\GT}\left( \vv_2^{(i)}/\sigma + L \vect{u}^{(i)}\right)$\\
		$\vv_2^{(i+1)} = \rho\tilde{\vv}_2^{(i+1)}+(1-\rho)\vv_2^{(i)}$\\
		\vspace{0.1em}
		\tcc{\scriptsize update corresponding to $h_3$}
		\vspace{0.1em}
		$\tilde{\vv}_3^{(i+1)} = \vv_3^{(i)} + \sigma K \vect{u}^{(i)} - \sigma\proj_{\GTF}\left( \vv_3^{(i)}/\sigma + K \vect{u}^{(i)}\right)$\\
		$\vv_3^{(i+1)} = \rho\tilde{\vv}_3^{(i+1)}+(1-\rho)\vv_3^{(i)}$\\
		\vspace{0.1em}
		\tcc{\scriptsize update of $\vect{u}$}
		\vspace{0.1em}
		$\vect{u}^{(i+1)} = \vect{u}^{(i)} - \rho\tau \adjoint{K}\adjoint{W}\left( 2\tilde{\vv}_1^{(i+1)} - \vv_1^{(i)} \right) - \rho\tau \adjoint{L}\left( 2\tilde{\vv}_2^{(i+1)} - \vv_2^{(i)}\right) - \rho\tau \adjoint{K}\left( 2\tilde{\vv}_3^{(i+1)} - \vv_3^{(i)}\right)$
	}
	\KwOut{$\vect{u}^{(i+1)}$}
	\caption{The CV algorithm
		%no.\ 2
		for solving the general formulation \eqref{eq:general.formulation}
		%, \eqref{eq:general.formulation.unconstrained}
	}
	\label{alg:CV}	
\end{algorithm}

\begin{algorithm}[t]
	\footnotesize
	\KwIn{The linear operators $W,K,L$, the proximal operator $\prox_\sparse$ and the projectors $\proj_{\GT},\proj_{\GTF}$.}
	Choose the parameters $\tau,\sigma,\rho>0$ satisfying the conditions \eqref{eq:CV.convergence}.\\ Choose the initial estimates $\vect{u}^{(0)},\vv_1^{(0)},\vv_2^{(0)}$.\\
	\For{$i = 0,1,\dots$}{
		\tcc{\scriptsize update corresponding to $h_1$}
		\vspace{0.1em}
		$\tilde{\vv}_1^{(i+1)} = \vv_1^{(i)} + \sigma WK \vect{u}^{(i)} - \sigma\prox_{\sparse/\sigma}\left( \vv_1^{(i)}/\sigma + WK \vect{u}^{(i)}\right)$\\
		$\vv_1^{(i+1)} = \rho\tilde{\vv}_1^{(i+1)}+(1-\rho)\vv_1^{(i)}$\\
		\vspace{0.1em}
		\tcc{\scriptsize update corresponding to $h_2$}
		\vspace{0.1em}
		$\tilde{\vv}_2^{(i+1)} = \vv_2^{(i)} + \sigma K \vect{u}^{(i)} - \sigma\proj_{\GTF}\left( \vv_2^{(i)}/\sigma + K \vect{u}^{(i)}\right)$\\
		$\vv_2^{(i+1)} = \rho\tilde{\vv}_2^{(i+1)}+(1-\rho)\vv_2^{(i)}$\\
		\vspace{0.1em}
		\tcc{\scriptsize notation for better readibility}
		\vspace{0.1em}
		$\w = \vect{u}^{(i)} - \tau \adjoint{K}\adjoint{W}\left( 2\tilde{\vv}_1^{(i+1)} - \vv_1^{(i)} \right) - \tau\adjoint{K}\left( 2\tilde{\vv}_2^{(i+1)} - \vv_2^{(i)} \right)$\\
		\vspace{0.1em}
		\tcc{\scriptsize update of $\vect{u}$}
		\vspace{0.1em}
		$\tilde{\vect{u}}^{(i+1)} = \w + \adjoint{L}\left( \proj_{\GT}(L\w) - L\w \right) $\\
		$\vect{u}^{(i+1)} = \rho\tilde{\vect{u}}^{(i+1)}+(1-\rho)\vect{u}^{(i)}$\\
	}
	\KwOut{$\vect{u}^{(i+1)}$}
	\caption{The CV algorithm
		%no.\ 2
		for solving the general formulation \eqref{eq:general.formulation}, assuming the use of a tight frame
		%, \eqref{eq:general.formulation.unconstrained}
		% \todo{Možná by bylo fajn mít Alg. 2 a 3 na stejné stránce}
	}
	\label{alg:CV.g}	
\end{algorithm}

Assuming a finite-dimensional problem together with $f = 0$,
the sequence $(\vect{u}^{(i)})_{i\in\NN}$ produced by the algorithm
is guaranteed to converge to the solution of problem \eqref{eq:CV.problem} if
\begin{equation}
	\tau\sigma\norm{\sum_{m=1}^{M}\adjoint{L_m} L_m} \leq 1,\quad 0 < \rho < 2.%
	\label{eq:CV.convergence}%
\end{equation}%
\pagebreak

To develop the case-specific form of Alg.\,\ref{alg:CV.general}, the functions from the formulation \eqref{eq:general.formulation.unconstrained} are assigned as follows:
\begin{subequations}
	\begin{align}
	h_1 &= \sparse, & h_2 &= \iota_{\GT}, & h_3 &= \iota_{\GTF},
	\label{eq:assignment:fc}\\
	L_1 &= W\!K, & L_2 &= L, & L_3 &= K,
	\label{eq:assignment:op}
	\end{align}%
	\label{eq:assignment}%
\end{subequations}%
and the functions $f,g$\/ are both zero.
Finally, we leverage the following general properties:
\begin{itemize}
	\item Since $g = 0$, it holds $\prox_{\tau g} = \Id$.
	\item To evaluate $\prox_{\sigma \adjoint{h}}$, where $\adjoint{h}$ is the Fenchel--Rockafellar conjugate of $h$, we use the Moreau identity 
	$\prox_{\sigma \adjoint{h}}(\vect{u}) = \vect{u} - \sigma\prox_{h/\sigma}(\vect{u}/\sigma)$ \cite{Moreau1965:Proximite.dualite}.
	\item The proximal operator of an indicator function $\iota_\Gamma$ of a closed convex set $\Gamma$ is the projection onto the set, denoted $\proj_\Gamma$.
\end{itemize}
%
%Although we are not aware of any general formula for $\prox_{\gamma h}$ once $\prox_h$ is known,
%the proximal operators
%$\prox_{\gamma\sparse} = \prox_{\gamma\norm{\cdot}_1}$, $\prox_{\gamma\iota_{\GT}}$, $\prox_{\gamma\iota_{\GTF}}$ are available for an arbitrary constant $\gamma>0$ in our setting.
%
Plugging these properties into Alg.\,\ref{alg:CV.general} produces the algorithm for the formulation \eqref{eq:general.formulation.unconstrained},
and thus for \eqref{eq:general.formulation}. 
The final algorithm is summarized in Alg.\,\ref{alg:CV}.
If the $\ell_1$ norm is used as the sparsity-inducing regularizer $S$, then $\prox_{\sparse/\sigma}$ becomes the soft thresholding.% operator.

The strength of the algorithm is that both projections can be performed explicitly and fast, entry by entry.
For the time-domain projection $\proj_{\GT}$, Eq.\,\eqref{eq:projT} is used.
For the TF-domain projection $\proj_{\GTF}$, the same equation can be adapted, since the projection can be done not only entrywise but also separately for the real and imaginary parts.
%\todo{Je to správně tak říct? Jo.}

Note that the functions in formulation \eqref{eq:general.formulation.unconstrained} were assigned to the functions $h_1, h_2, h_3$
such that Alg.\,\ref{alg:CV} covers both the synthesis and the analysis approaches \eqref{eq:general.formulation.synthesis} and \eqref{eq:general.formulation.analysis}, respectively.
Had the composition $\sparse\circ (W\!K)$ been assigned to the function $g$ instead, the operator $\prox_{\tau g}$ would be known only in the synthesis model.\!%
\footnote{The potential evaluation of $\prox_{\tau g} = \prox_{\tau \sparse\circ\ana}$ in the analysis model is complicated,
because the formula for a proximal operator of such a composition is known only when the  operator $\ana$ satisfies $\ana\syn=\alpha\Id$, which is not possible in the setting of redundant TF transforms
\cite{RajmicZaviskaVeselyMokry2019:Axioms}.}

\subsection{Consistent convex approach, tight frame case}
\label{sec:consistent.convex.2}

%As mentioned in \cite{Condat2014:Generic.proximal.algorithm}, if possible,
%one should make use of the functions $f$ and $g$ in Eq.\,\eqref{eq:CV.problem} when assigning the functions of a particular problem.
Alternatively, we can make the assignment such that the function $g$ is used.
In \cite{Condat2014:Generic.proximal.algorithm},
it is suggested that employing the function $g$ may result in a faster convergence of the algorithm.
Such an assignment is not possible in the case of the formulation \eqref{eq:general.formulation.unconstrained},
unless the linear operators represent the analysis or synthesis of a~tight frame.
In such a~special case, we may assign
\begin{subequations}
	\begin{align}
	g &= \iota_{\GT}\circ L, &h_1 &= \sparse, & h_2 &= \iota_{\GTF},
	\label{eq:assignment_g:fc}\\
	&& L_1 &= W\!K, & L_2 &= K.
	\label{eq:assignment_g:op}
	\end{align}
	\label{eq:assignment_g}%
\end{subequations}%
This is justified by the observation that in the case of a tight frame,
$L$ is either the synthesis (in the synthesis model),
or identity (in the analysis model).
In both cases, it satisfies $L\adjoint{L} = \alpha\Id$ for a~positive constant $\alpha$,
allowing us to compute the proximal operator $\prox_{\iota_{\GT}\circ L}$ using the explicit formula \cite{combettes2011proximal,RajmicZaviskaVeselyMokry2019:Axioms}
\begin{equation}
	\prox_{\iota_{\GT}\circ L}(\vect{u}) = \vect{u} + \alpha^{-1}\adjoint{L}\left( \proj_{\GT}(L\vect{u}) - L\vect{u} \right).
\end{equation}
Put in words, the formula states that instead of computing the complicated projection on the left-hand side,
one may use the simple projection onto $\GT$ on the right-hand side,
together with the application of the linear operator and its adjoint.

The resulting algorithm is summarized by Alg.\,\ref{alg:CV.g},
where, for simplicity, $\alpha = 1$ is assumed (i.e.\ the frame is Parseval tight).
Compared to Alg.\,\ref{alg:CV}, this algorithm has a~major benefit:
%\begin{itemize}
	%\item for $\rho\leq 1$, every iterate $\vect{u}^{(i+1)}$ lies in $\GT$, %the \todo{solution} in every iteration of the algorithm lies in $\GT$, \todo{dokázat?}
	%\todo{proč je to benefit, Ondro? \textbf{Protože jsme konzistentní v každé iteraci, tzn. i při malém počtu iterací.}}
	%\item in \cite{Condat2014:Generic.proximal.algorithm},
	%it is suggested that involving the function $g$ may result in faster convergence of the algorithm,
	%\item 
	since it uses only two functions $h_1,h_2$ and thus only two corresponding linear operators,
	it follows from Eq.\,\eqref{eq:CV.convergence} that a~wider range of the parameters $\tau,\sigma$ is allowed,
	creating the possibility for faster convergence.

\subsection{Inconsistent convex approach}
\label{sec:inconsistent.convex}

So far, the solutions to all of the reconstruction tasks have been assumed to be consistent with the observations
(either time-domain samples or TF coefficients, or even both).
%\todo{i.e.\ a strict requirement to belong to Gamma...}.
However, this assumption may be too strong, for example in the case of noisy data.
In such a~case, instead of strictly forcing the signal to lie in $\GT$ and the coefficients to lie in $\GTF$, we minimize the distances to these sets.
The formulation \eqref{eq:general.formulation.unconstrained} would cover also this case,
had we used the distance from $\GT$ and $\GTF$ instead of the indicator functions 
(which force the respective distance to be zero).

Since the proximal operator of a~distance function of a closed convex set is available \cite{combettes2011proximal},
the inconsistent problem could be solved by the CV algorithm, similarly to the consistent one in Sec.\,\ref{sec:consistent.convex.1} or \ref{sec:consistent.convex.2}.

%\subsection{Inconsistent non-convex approach}
%\label{sec:inconsistent.nonconvex}
%
%Similarly to the previous approach, the inconsistent non-convex approach is naturally obtained by modifying the consistent one.
%As mentioned above, the consistent SPARE algorithm would involve a~projection onto $\Gone\cap\Gtwo$ in each iteration, ensuring the consistency of the resulting signal.
%Relaxing this step such that it corresponds to the proximal operator of distance from the set $\Gone\cap\Gtwo$ directly produces the inconsistent variant of SPARE.

\section{Experiment}
\label{sec:experiment}
We perform an experiment that serves as the proof of concept of the presented recovery formulation.
On top of that, the results suggest interesting implications that could lead to new developments in audio coding;
we show that a simultaneous utilization of the time and time-frequency information could lead to better compression in some cases, compared to conventional, single-domain approaches.
%\todo{to je možná příliš silné tvrzení, protože single TF-domain jsme neprováděli?}

\begin{figure}
	\centering
	\usetikzlibrary{backgrounds,calc,positioning}
\definecolor{mygray}{gray}{0.6}

\scalebox{0.9}{%
\begin{tikzpicture}[every text node part/.style={align=center}]

\node[draw=mygray, thick, minimum width={width("coefs $\in C^Q$")+1pt}] 
	(v1) at (-3.5,2.5) {signal $\in \RR^P$};

\node[thick, minimum width={width("coefs $\in C^Q$")+1pt}] 
	(v4) at (-3.5,1.2) {coefs $\in \CC^Q$};

\node[thick, minimum width={width("coefs $\in C^{pTF\cdot Q}$")+6pt}] 
	(v2) at (0,2.5) {subsampled \\ signal $\in \RR^{\pT\cdot P}$};

\node[thick, minimum width={width("coefs $\in C^{pTF\cdot Q}$")+6pt}] 
  (v5) at (0,1.2) {subsampled \\ coefs $\in \CC^{\pTF\cdot Q}$};

\node[thick, minimum width={width("$pTF\!\cdot\!Q\!\cdot\! bTF$ bits")+6pt}]
	(v3) at (3.5,2.5) {$\pT  \!\cdot\! P \!\cdot\! \bT$ bits};

\node[thick, minimum width={width("$pTF\!\cdot\!Q\!\cdot\! bTF$ bits")+6pt}] 
	(v6) at (3.5,1.2) {$\pTF\!\cdot\!Q\!\cdot\! \bTF$ bits};

\draw[-latex]  (v1) to node[midway, sloped, above]{\textcolor{mygray}{subs.}}  (v2);
\draw[-latex]  (v2) to node[midway, sloped, above]{\textcolor{mygray}{quant.}}  (v3);
\draw[-latex]  (v1) to node[midway, left]{\textcolor{mygray}{DGT}} (v4);
\draw[-latex]  (v4) to node[midway, sloped, above]{\textcolor{mygray}{hard}\\\textcolor{mygray}{thresh.}} (v5);
\draw[-latex]  (v5) to node[midway, sloped, above]{\textcolor{mygray}{quant.}} (v6);

\node[draw=mygray, thick, minimum width={width("$pTF\!\cdot\!Q\!\cdot\! bTF$ bits")+6pt}, minimum height = 1.85cm] at (3.5,1.85) {};

\end{tikzpicture}
}
	\vspace*{-9pt}
	\caption{Scheme of the degradation considered in the experiment.
	The abbreviation DGT stands for the discrete Gabor transform in place of the analysis operator $\ana$.}
	\label{fig:scheme}
\end{figure}
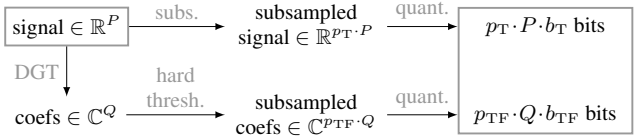

\begin{figure*}%[h]
	\centering
	\begin{subfigure}[b]{0.32\textwidth}
		\includegraphics[width=\textwidth]{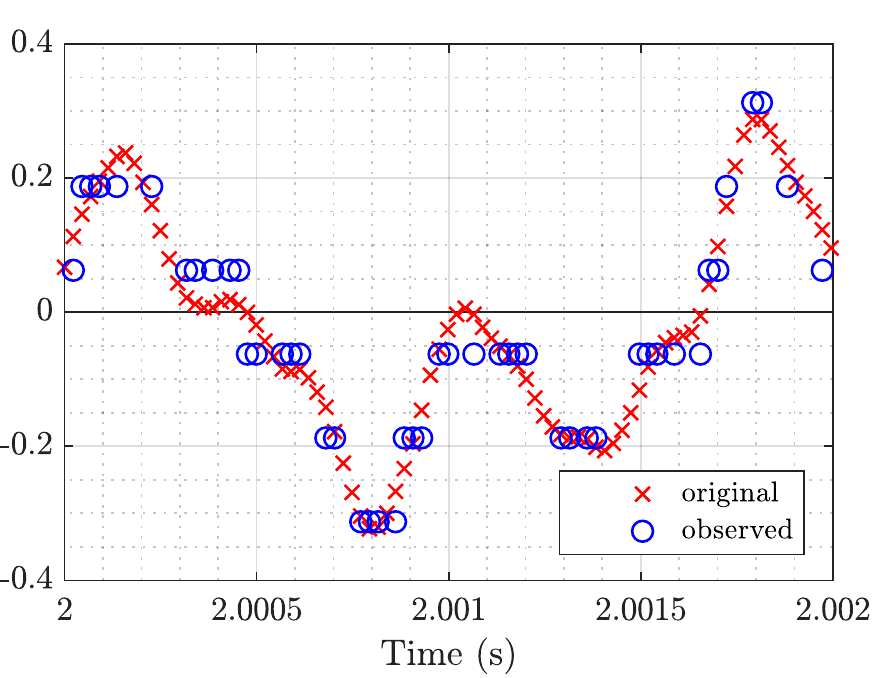}
		\caption{piece of the subsampled and quantized signal} %in the time domain}
		\label{fig:observation:T.domain.observed}
	\end{subfigure}
	\hfill
	\begin{subfigure}[b]{0.32\textwidth}
		\includegraphics[width=\textwidth]{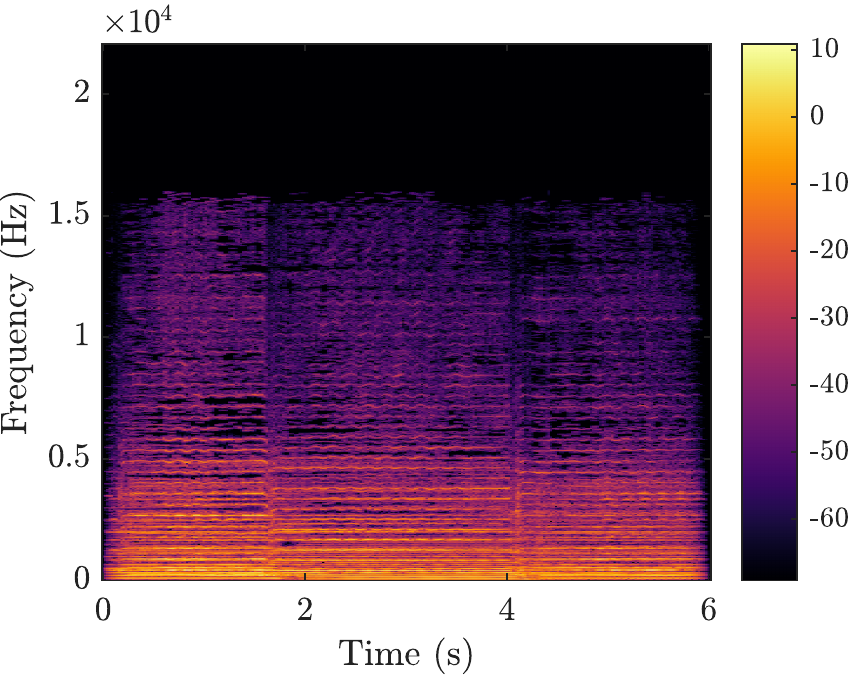}
		\caption{spectrogram of the original signal} %\\ \vspace{1em}}
		\label{fig:observation:TF.domain.original}
	\end{subfigure}
	\hfill
	\begin{subfigure}[b]{0.32\textwidth}
		\includegraphics[width=\textwidth]{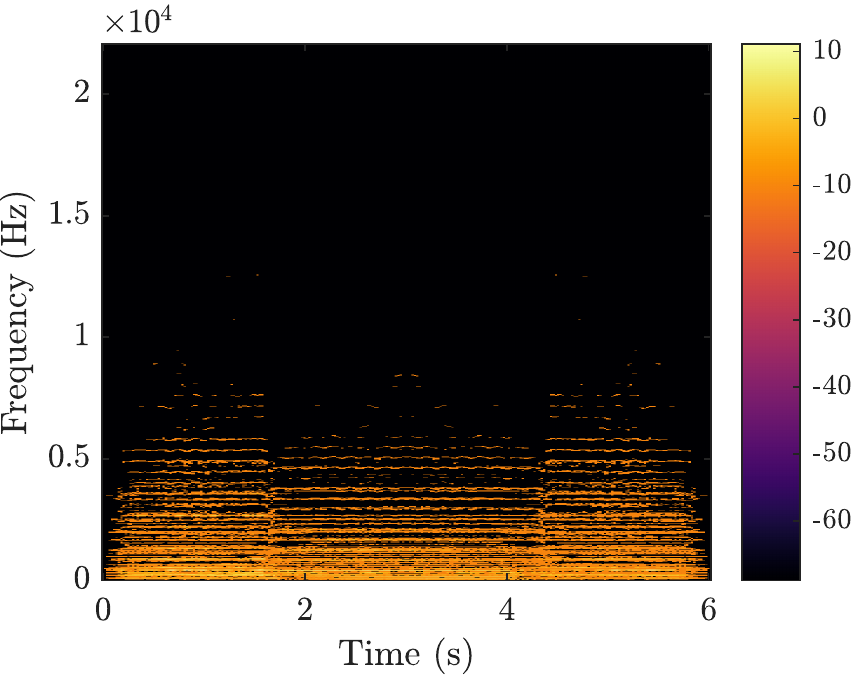}
		%\caption{observed spectrogram\\ \vspace{1em}}
		\caption{spectrogram with kept quantized values} %\\ \vspace{1em}}
		\label{fig:observation:TF.domain.observed}
	\end{subfigure}
	\caption{Data available to the decoder:
	(a) subsampled and quantized time-domain samples, and
	(c) subsampled and quantized TF coefficients.
	Although the real and imaginary parts are treated separately in the TF domain, the magnitude spectrogram is depicted here.
	}%
	\label{fig:observation}
\end{figure*}

\subsection{Design of the experiment}
The task is to reconstruct a~signal where some samples are missing;
moreover, the retained samples are  quantized.
At the same time, a~partial and quantized observation of the TF coefficients of the original (non-distorted) signal is provided.
The goal is to illustrate that it is beneficial to utilize the double-domain approach,
compared to the \restoration{} using only information in the time domain (abbreviated to T domain in some of the figures).
The relaxed sparse prior, i.e.\ the $\ell_1$ norm, is used,
hence we can apply the consistent convex approach from Sec.\,\ref{sec:consistent.convex.2}.

The percentage of available samples/coefficients varies from $10\,\%$ up to $90\,\%$.
It is denoted by $\pT$ and $\pTF$, respectively.
In the time domain, the reliable samples are distributed (uniformly) randomly.
In the TF domain, the coefficients that are the largest in magnitude are kept
(Sec.\,\ref{sec:complex.TF.coefficients} gives additional comments on the choice of the coefficients).
The quantization is uniform and it is done by
%virtually
limiting the number of bits per sample ($\bT$) or per coefficient ($\bTF$).
For a~given bit depth $b$ (i.e.\ the number of bits used for representing each number),
%, bps),   %...PR: vyhodil jsem, nikde se to pak nepoužije a navíc je to zaměnitelné s bits per second
$\Delta=2^{-b+1}$ denotes the distance of two consecutive quantization levels.
The quantized observation $u^\mathrm{quant}$ of a~real value $u$, $-1\leq u\leq 1$ is obtained using the so-called \textit{mid-riser uniform quantizer}
\cite{Gray1998:Quantization} as
\begin{equation}
	u^\mathrm{quant} = \operatorname{sgn}^{\!+}(u)\left( \left\lfloor \frac{\abs{u}}{\Delta} \right\rfloor +\frac{1}{2} \right),
	\label{eq:mid.riser}
\end{equation}
where $\operatorname{sgn}^{\!+}(u)$ returns 1 for $u\geq 0$ and $-1$ for $u<0$.
%In the present experiment,
The bit depths $\bT$ and $\bTF$
are chosen as the powers of two and they are equal, $\bT = \bTF\in\{2,4,8,16,32\}$.
The samples or coefficients considered lost are the only exception, they are simply set to zero. % independently of the quantization.}
%(the original non-distorted signal is saved in \qm{double-precision} floating point format in MATLAB, i.e.\ using 64 bits).
%\todo{Tady by to chtělo to kvantování dobře promyslet, i pro článek, ale hlavně pro nás do budoucna. V plovoucí řádové čárce je to s těmi čísly složitější (myslím). V MATLABu neexistuje floating point např. pro 8 bitů. Je potřeba si rozmyslet, jak to vlastně dělat (tj. Pavlova funkce) a jak to podat v textu.}

As the TF transform, the discrete Gabor transform (DGT) is used, with the sine window of 2048 samples in length, 50\,\% overlap and 2048 frequency channels.
Such a transform produces a twice-redundant tight frame, which is then normalized to obtain a~Parseval tight frame.
%Such a transform produces a tight frame ($\alpha=2$), which is then normalized to obtain a~Parseval tight frame.
As the prior, we use
$\sparse\!=\!\norm{\cdot}_1$ with no weighting, i.e.\ $W=\Id$.

For an illustrative scheme of the degradation, see Fig.\,\ref{fig:scheme}.
Fig.\,\ref{fig:observation} then shows an example of the degraded signal and coefficients.

In order to evaluate the results,
the PEMO-Q ODG score \cite{Huber:2006a} and the SDR are measured,
the latter being defined as
\begin{equation}
	\sdr(\y,\hat{\y}) = 10\log_{10}\frac{\norm{\y}^2}{\norm{\y-\hat{\y}}^2},
\end{equation}
where $\y$ is the original (non-distorted) time-domain signal and $\hat{\y}$ is the reconstruction.
The result is expressed in decibels.
Unlike the SDR, PEMO-Q is a~perceptually motivated measure whose ODG output score ranges from $-4$ (very annoying distortion, poor quality) to $0$ (imperceptible distortion, excellent quality).

The experiment is run for a set of 10 audio signals (musical recordings) of varying complexity from the SQAM database \cite{EBUSQAM}.
The signals are sampled at 44.1\,kHz. % and they are originally ca.\ 5\,s long.
To reduce the computational time due to the enormous number of tested combinations, the proof-of-concept experiment only uses one-second long excerpts.
A single \restoration{} instance then takes ca.\ 5\,s, depending on the parameters of the computer.
For the purpose of quantization, these excerpts are also peak-normalized such that the maximum absolute value of each signal equals one.

The CV algorithm \ref{alg:CV.g} is executed setting $\tau = \sigma = \sqrt{2}/2$, $\rho = 1$, and it stops after 300 iterations.
The choice of $\tau$ and $\sigma$ follows from \eqref{eq:CV.convergence} and \eqref{eq:assignment_g:op},
since $\norm{\sum_{m=1}^{2}\adjoint{L_m} L_m} = \norm{\adjoint{K}K + \adjoint{K}K} = 2\norm{\adjoint{K}K} = 2$
both in the synthesis case ($K=\Id$) and in the analysis case ($K=\ana$), assuming a Parseval tight frame and
%no weighting
$W=\Id$.

\subsubsection{On the choice and quantization of the TF coefficients}
\label{sec:complex.TF.coefficients}

In the experiment, a~tight Gabor frame is used to compute the TF representation of a real signal.
Coefficients obtained using such a~frame attain a~specific complex-conjugate structure.
In fact, only a~half of all the coefficients are needed; the other half may be computed as a conjugate to the first half.
Such a structure introduces a~kind of redundancy:
A pair of coefficients, given they are complex-conjugate, contribute to the total bit rate by the same amount as a pair of real samples of the signal.
This property is used in the implementation when choosing the subset of the TF coefficients;
it is ensured that for a~given number of reliable samples or coefficients,
information from the TF domain yields the same bit rate as information from the time domain.

Furthermore, recall that the quantization defined by Eq.\,\eqref{eq:mid.riser} is tailored for values from the interval $[-1,1]$.
%\todo{což je vhodné pro T doménu; normalizuješ na úplném začátku ty signály, BTW?}
To simulate the quantization for the observed TF coefficients $\vect{c}$, the quantization step $\Delta$ and all the quantization and decision levels in the TF domain are scaled by a factor of $\max\{\max\{\abs{\Re(\vect{c})}\},\;\max\{\abs{\Im(\vect{c})}\}\}$.

\subsection{Results}

\begin{figure*}[ht]
	\centering
	\begin{subfigure}[b]{0.31\textwidth}
		\includegraphics[width=\textwidth]{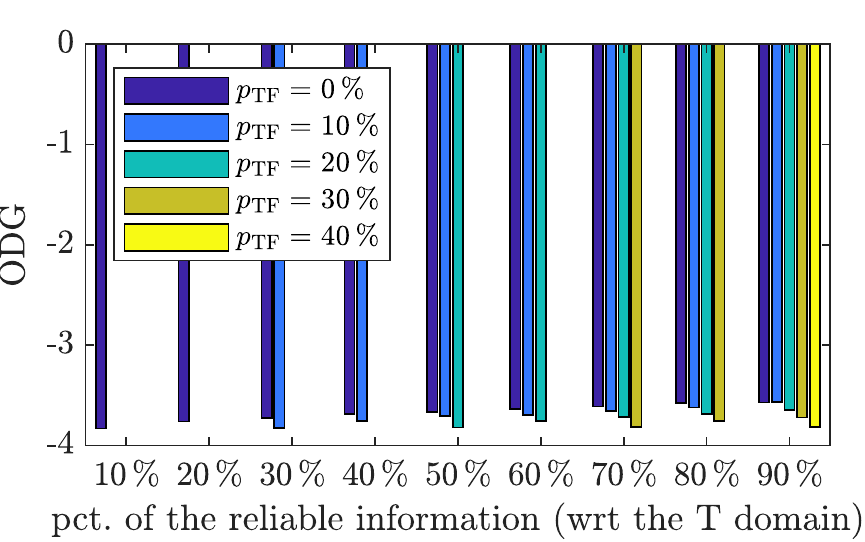}
		\caption{ODG, synthesis model, 4 bits}
		\label{fig:bar:ODGsyn4bit}
	\end{subfigure}
	\hfill
	\begin{subfigure}[b]{0.31\textwidth}
		\includegraphics[width=\textwidth]{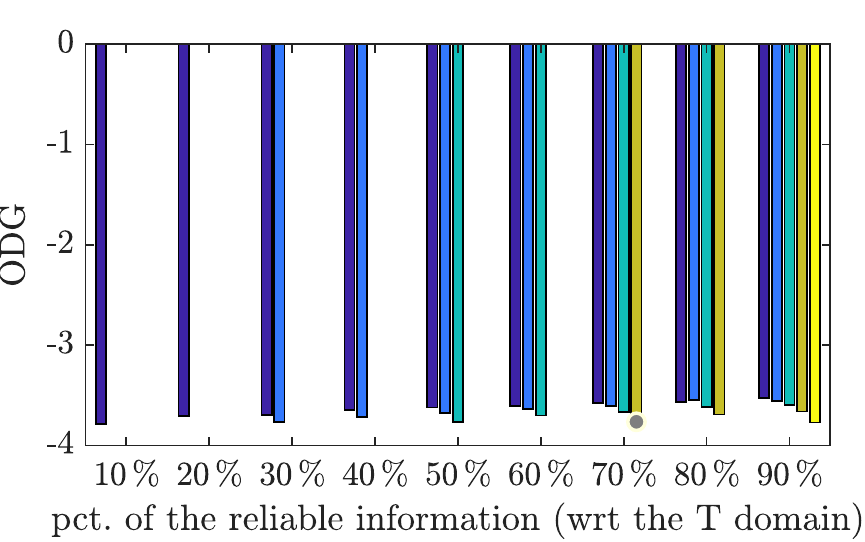}
		\caption{ODG, analysis model, 4 bits}
		\label{fig:bar:ODGana4bit}
	\end{subfigure}
	\hfill
	\begin{subfigure}[b]{0.31\textwidth}
		\includegraphics[width=\textwidth]{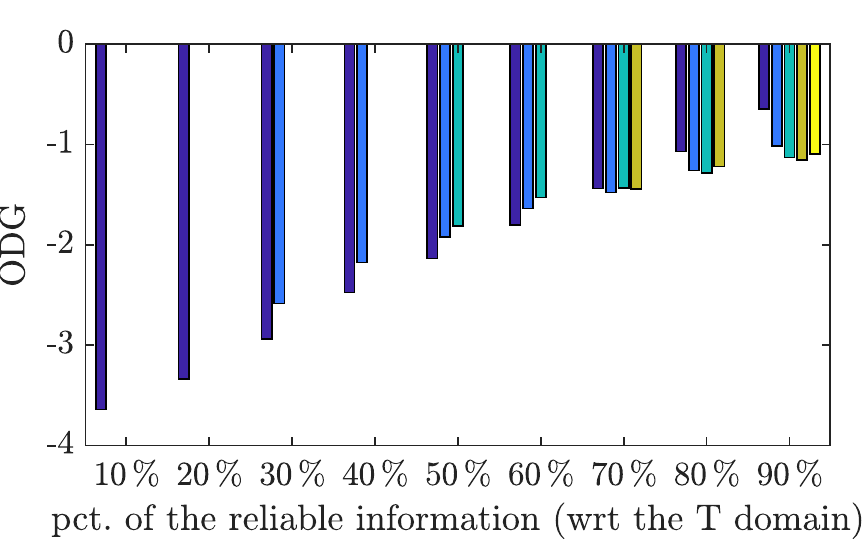}
		\caption{ODG, analysis model, 16 bits}
		\label{fig:bar:ODGana16bit}
	\end{subfigure}\\
	\vspace{-0.3em}
	\begin{subfigure}[b]{0.31\textwidth}
		\includegraphics[width=\textwidth]{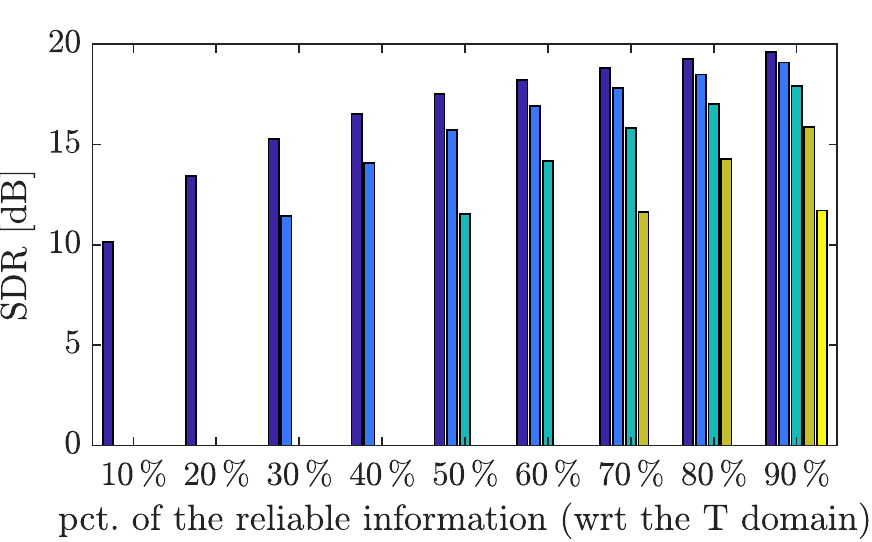}
		\caption{SDR, synthesis model, 4 bits}
		\label{fig:bar:SDRsyn4bit}
	\end{subfigure}
	\hfill
	\begin{subfigure}[b]{0.31\textwidth}
		\includegraphics[width=\textwidth]{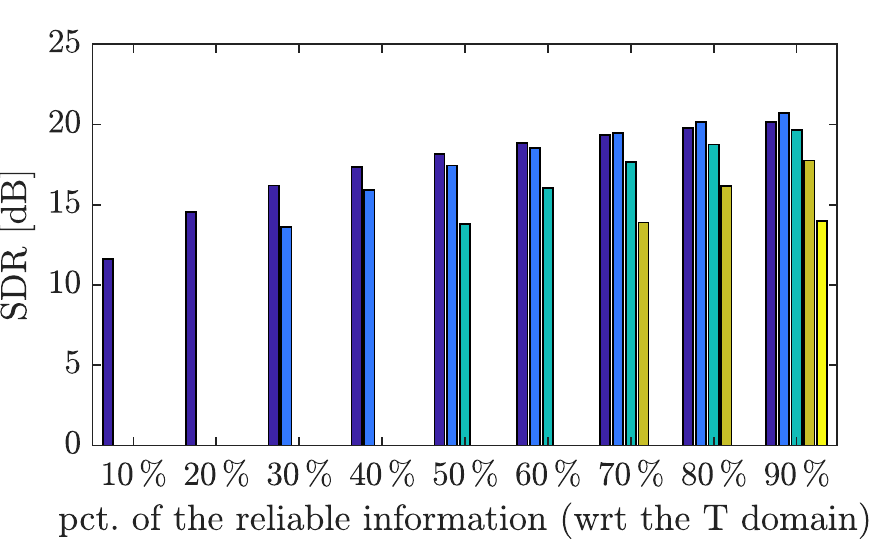}
		\caption{SDR, analysis model, 4 bits}
		\label{fig:bar:SDRana4bit}
	\end{subfigure}
	\hfill	
	\begin{subfigure}[b]{0.31\textwidth}
		\includegraphics[width=\textwidth]{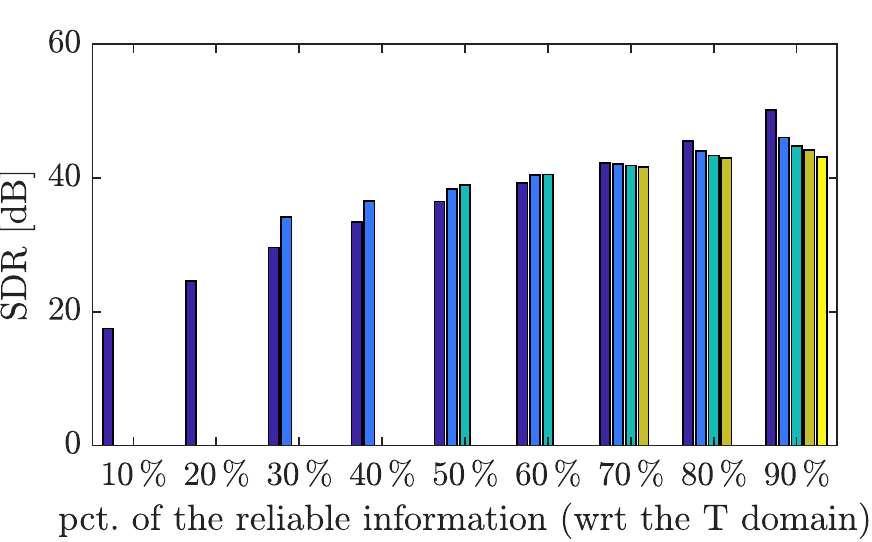}
		\caption{SDR, analysis model, 16 bits}
		\label{fig:bar:SDRana16bit}
	\end{subfigure}
	\vspace{-0.5em}
	\caption{Comparison with fixed bit depth. The legend shown in the first plot is common to all the plots;
	%The notation
	$\pTF$ denotes the percentage of reliable TF coefficients.
	Here, reliable coefficient means that it is observed (i.e.\ not missing), although it is quantized.}
	\label{fig:bar}
	\vspace{-1em}
\end{figure*}

\begin{figure*}[h!]
	\centering
	\begin{subfigure}[b]{0.43\textwidth}%47
		\includegraphics[width=\textwidth]{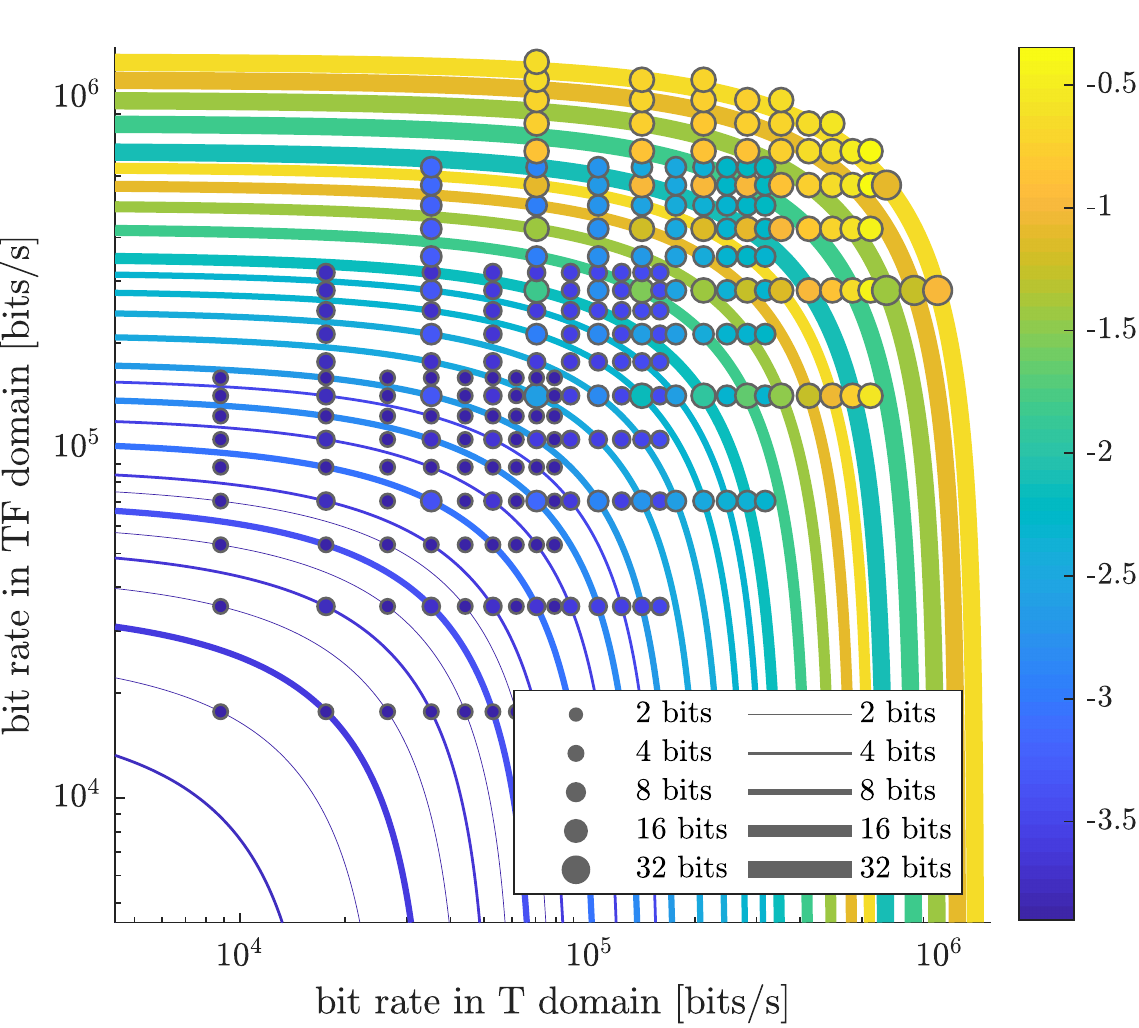}
		\caption{ODG, analysis model}
		\label{fig:scatter:ODG}
	\end{subfigure}
	\hspace{0.05\textwidth}
	\begin{subfigure}[b]{0.43\textwidth}
		\includegraphics[width=\textwidth]{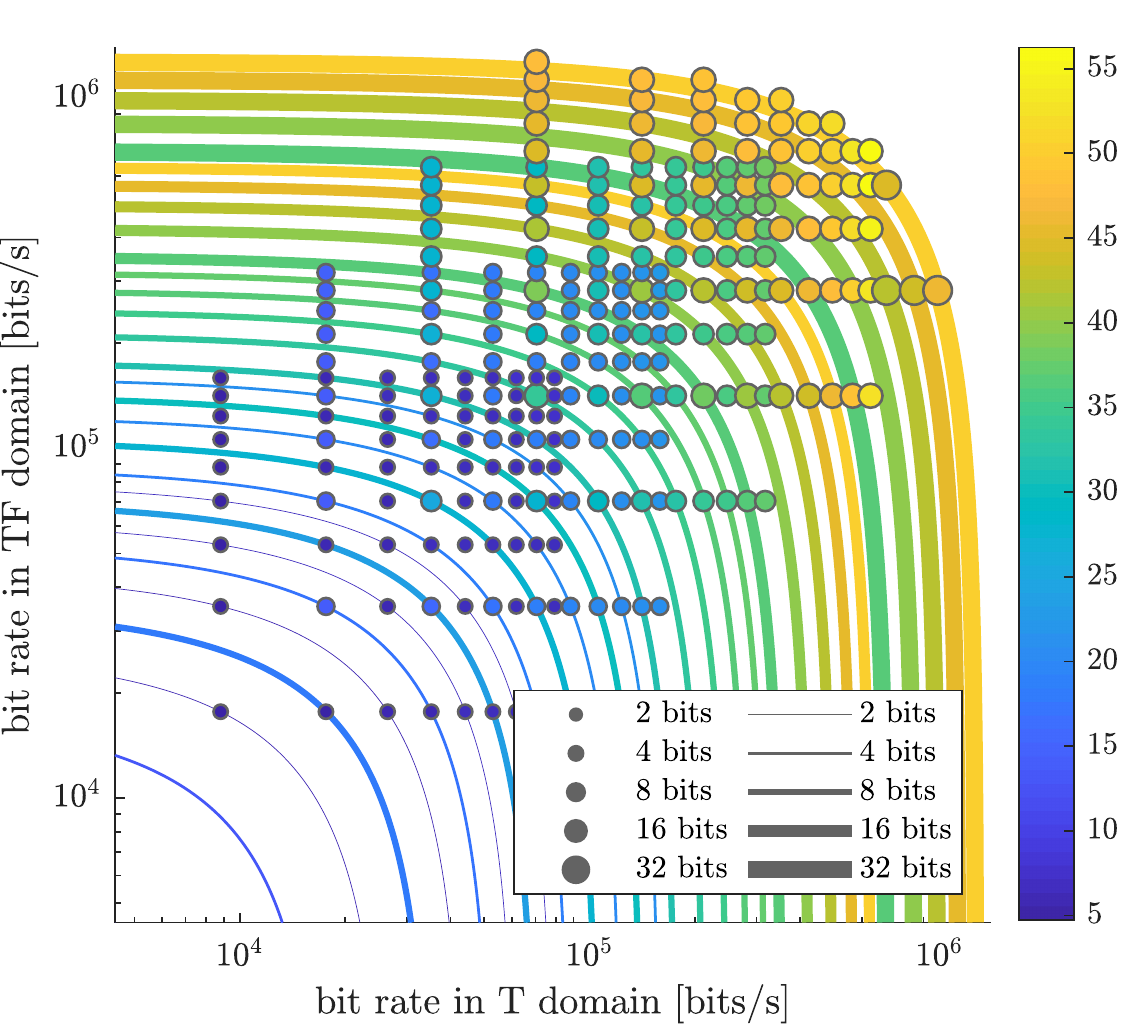}
		\caption{SDR, analysis model}
		\label{fig:scatter:SDR}
	\end{subfigure}
	\vspace{-0.5em}
	\caption{Comparison with variable bit depth. The bit rate is the quantity of bits per a~second of audio.}
	\label{fig:scatter}
	\vspace{-1em} %-0.33
\end{figure*}

%\begin{figure}[h!]
%	\centering
%	\begin{subfigure}[b]{0.73\linewidth}
%		\includegraphics[width=\textwidth]{figures/lineODGana.pdf}
%		\caption{ODG, analysis model}
%		\label{fig:line:ODG}
%	\end{subfigure}
%	\\
%	\begin{subfigure}[b]{0.73\linewidth}
%		\includegraphics[width=\textwidth]{figures/lineSDRana.pdf}
%		\caption{SDR, analysis model}
%		\label{fig:line:SDR}
%	\end{subfigure}
%	%
%	\caption{Comparison of the best performance of different approaches, given the limit of available total bit rate.}
%	%
%	\label{fig:line}
%\end{figure}

\begin{figure*}[h!]
	\centering
	\begin{subfigure}[b]{0.43\linewidth}
		\includegraphics[width=\textwidth]{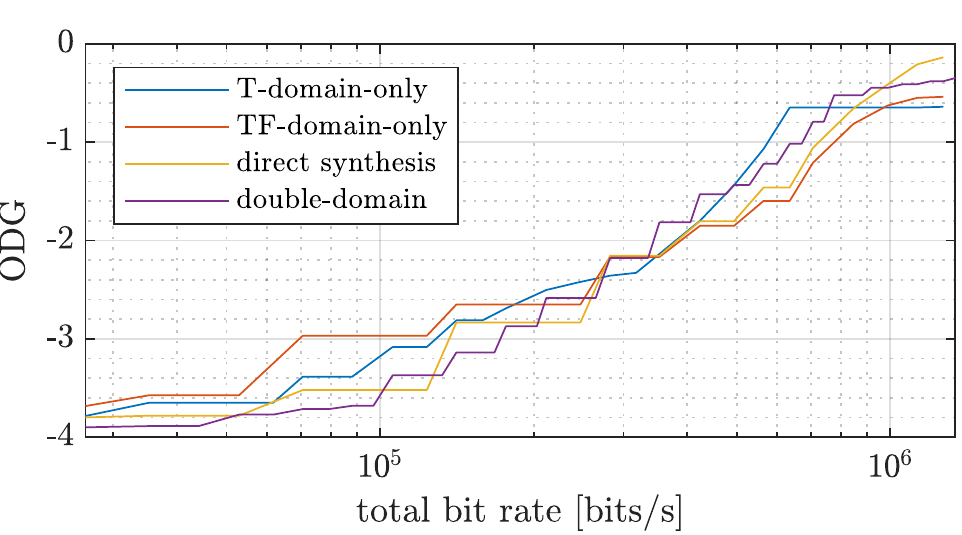}
		\caption{ODG, analysis model}
		\label{fig:line:ODG}
	\end{subfigure}
	\hspace{0.05\textwidth}
	\begin{subfigure}[b]{0.43\linewidth}
		\includegraphics[width=\textwidth]{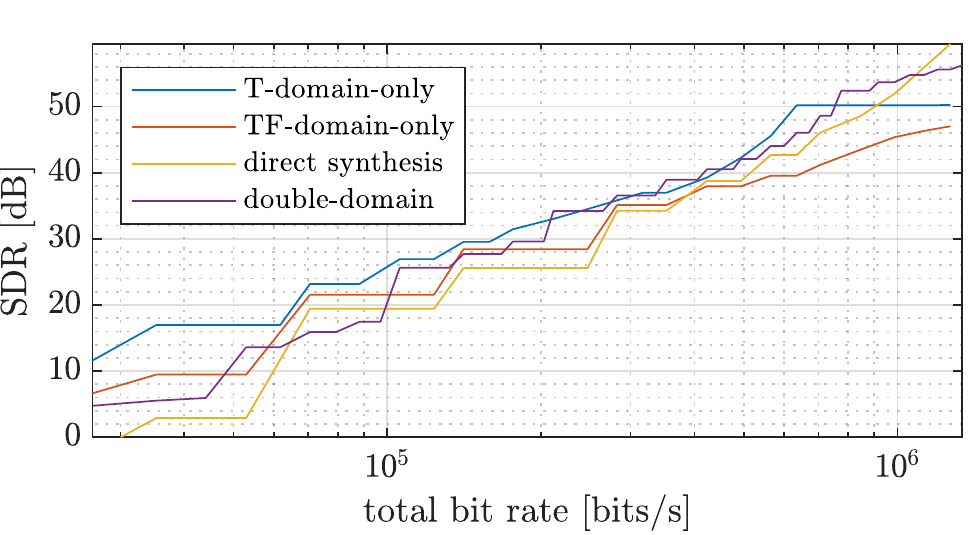}
		\caption{SDR, analysis model}
		\label{fig:line:SDR}
	\end{subfigure}
	\vspace{-0.5em}
	\caption{Comparison of the best performance of different approaches, given the limit of available total bit rate.}
	\label{fig:line}
	%\vspace{-1em}
\end{figure*}

\subsubsection{Comparison with fixed bit depth}

All the results are visualized as mean values computed from the 10 audio signals.
In the first visualization in Figure \ref{fig:bar}, the bit depth is fixed.
The result corresponding to the T-domain-only approach (denoted by $\pTF = 0\,\%$) with a~given fraction of reliable samples in the time domain serves as a reference.
These two parameters---bit depth and fraction---define the bit rate of reliable information used in the T-domain-only approach.
This reference scenario is compared to different distributions of the total amount of bits 
between the time and TF domains while using the previously fixed bit depth.
Note that only a limited number of options of how to distribute the information between the time and the TF domains was tested.

%Figure \ref{fig:bar} shows the results for a single audio excerpt.
Both evaluation metrics (ODG and SDR) are depicted in Fig.\,\ref{fig:bar}.
For the bit depth $\bT=\bTF=4$, we present the results using both the analysis and the synthesis models
(plots \ref{fig:bar:ODGsyn4bit}, \ref{fig:bar:ODGana4bit}, \ref{fig:bar:SDRsyn4bit}, \ref{fig:bar:SDRana4bit}).
Since no significant difference between the performance of the analysis and the synthesis approaches is observed, only the analysis model is used for further comparison with the performance using $\bT=\bTF=16$
(plots \ref{fig:bar:ODGana16bit} and \ref{fig:bar:SDRana16bit}).

For a~fixed number of bits per sample or coefficient, it is in general not beneficial to split the available information between the two domains;
see the decrease in both ODG and SDR in the plots \ref{fig:bar:ODGsyn4bit}, \ref{fig:bar:ODGana4bit}, \ref{fig:bar:SDRsyn4bit} and \ref{fig:bar:SDRana4bit} when the percentage of reliable TF coefficients increases.
Sampling in the TF domain (in our setup)
is reasonable only with a high bit depth---compare, for example, plots
\ref{fig:bar:SDRana4bit} and \ref{fig:bar:SDRana16bit}, where the difference is less significant.

\subsubsection{Comparison with variable bit depth}

In the visualization in Fig.\,\ref{fig:scatter},
the number of bits per sample or coefficient varies.
Two ways of displaying the results are combined in the figure.

The T-domain-only approach is represented by the colored equibital lines%
\footnote{i.e. lines connecting points with the same total bit rate}\!.
The line color represents the restoration quality, according to the side colorbar.
The line width represents the bit depth and the position represents the bit rate
(in this case, only time-domain information is used).

The double-domain approach is represented by the colored points.
Once again, the color indicates the restoration quality.
The point size represents the bit depth $\bT=\bTF$.
Finally, the position represents the distribution of reliable information between the domains.
%i.e.\ points lying under the 45-degree line correspond to experiments where a~greater bit rate has been assigned to the time domain.

Both in the case of lines and in the case of points, the following rule is applied:
If more realizations with the same bit distribution appear, only the best of them is plotted.
Such a situation occurs when we decrease the number of reliable samples/coefficients while we increase the bit depth.

%\subsubsection{Discussion on the results}
%\todo{Možná přemístit interpretace přímo k těm dvěma částem.}

The scatter plots in Figure \ref{fig:scatter} show that there is a number of cases where it is useful to decrease the precision of the reliable time-domain samples and
assign a~part of the bit budget to the TF domain.
This conclusion can be deduced from points which lie on an equibital line.
%(i.e. the same total bit rate is used for the reliable information).
%If the color of the point indicates that the restoration quality is higher compared to the one indicated by the color of the line, it means that using the information in the TF domain instead of the time domain is beneficial.
Using the TF-domain information is advantageous when such a~point reports a higher ODG/SDR compared to the line it lies on.

The evaluation is concluded by Fig.\,\ref{fig:line}, which provides a different perspective to Fig.\,\ref{fig:scatter}.
The figure considers only the best performance in terms of ODG (Fig.\,\ref{fig:line:ODG}) or SDR (Fig.\,\ref{fig:line:SDR}) as a function of available total bit rate.
In other words, it does not consider results in a situation when a higher bit rate does not lead to a better performance (cf. Fig.\,\ref{fig:scatter:ODG}, the 5th and the 6th equibital from the top).
The plots show that there are cases where the double-domain approach outperforms the T-domain-only approach, however, the difference is only minor.
Significant gain is observed only for the highest bit rates, i.e.\ a low level of compression.

For the sake of completeness, the TF-domain-only reconstruction is included in Fig.\,\ref{fig:line} as well.
%Similarly to the T-domain-only approach, only the information from one domain is available to the decoder.
In this case, the reconstruction is carried out based solely on partially observed and quantized TF coefficients.
We consider two options:
Either we follow the framework where the observation induces the set $\GTF$, and Alg.\,\ref{alg:CV.g} provides the reconstruction, or the quantized, partially observed coefficients are directly synthesized with $\syn$.
Interestingly, both the T-domain-only and the double-domain approaches remain superior in great number cases; in particular regarding the SDR. %to form the resulting signal.
%The second option naturally %simple idea to synthesize the observed samples with no prior clearly reproduces the errors induced by the quantization.
%Thus, the method 
%does not perform well for low bit rates (harsh quantization), but outperforms the rest of the methods for the highest bit rates (minor quantization, high percentage of reliable coefficients).
%Similar results are observed for the TF-domain-only approach; the optimization procedure only flattens the extremes of the curve.

\subsection{Software and reproducible research}
\label{sec:software}
The experiment was run in MATLAB R2019b, using \mbox{LTFAT} \cite{LTFAT} version 2.3.1.
All MATLAB codes, together with supp\-le\-men\-tal ma\-te\-ri\-al, are provided in the repository at \url{https://github.com/ondrejmokry/AudioRestorationFramework/}.

\section{Conclusion}
The paper provides a~general flexible formulation not only co\-ve\-ring multiple audio \restoration{} tasks,
but also allowing several degradation types to take place simultaneously.
Another novelty is that the restoration can possibly take into account constraints in the time-frequency domain.
The concept can be easily extended such that the reliable information is distributed among more than two different transform domains.
In Sec.\,\ref{sec:inconsistent.convex}, it is proposed how to develop the framework such that it handles noise-distorted data.

The aim of the experiment was not to outperform the state-of-the-art methods in the field of audio \restoration, but to show an application of the general formulation in a~meaningful audio compression scenario.
The framework was shown to be flexible enough to cover a model of signal distortion
which included both drop-outs and quantization of both the samples in the time domain and of the time-frequency coefficients.
Although only a single scenario was considered, we observed promising results.
It remains for the future work to find an optimal distribution of the bit budget between the time and the time-frequency domains.
%Even such a brief example demonstrates that it is worth studying possible distributions of reliable information.
%, since it may yield remarkable results.

\section{Acknowledgments}

The work was supported by the Czech Science Foundation (GA\v{C}R) project number 20-29009S.
The authors thank the reviewers for their comments and suggestions, especially the reviewer number~4.

%\newpage
%\nocite{*}
	
%\inputencoding{cp1250}
%\bibliographystyle{IEEEbib}
%\bibliography{literatura}

\newcommand{\noopsort}[1]{} \newcommand{\printfirst}[2]{#1}
\newcommand{\singleletter}[1]{#1} \newcommand{\switchargs}[2]{#2#1}

\end{document}